\documentclass[aps,pra,reprint]{revtex4-2}

\usepackage{amsmath}
\usepackage{amssymb}
\usepackage{graphicx}
\usepackage{hyperref}
\usepackage{blindtext}

\begin{abstract}
Radially-polarized light beams present very interesting and useful behavior for creating small intensity spots when tightly-focused, and manipulating nanostructures or charged particles. The modeling of the propagation of such vector beams, however, is almost always done using the lowest-order fundamental radially-polarized beam due to the complexity of vector diffraction theory. We show how a flat-top radially-polarized beam can be modeled analytically using a sum of higher-order beams, and describe a number of interesting qualities, and compare to numerically-solved integral descriptions.
\end{abstract}
\begin{document}

\title{Modeling the focusing of a radially-polarized laser beam with an initially flat-top intensity profile}
\author{Spencer W. Jolly}
\email{spencer.jolly@ulb.be}
\affiliation{Service OPERA-Photonique, Université libre de Bruxelles (ULB), Brussels, Belgium}
\date{\today}
\maketitle

\section{Introduction}
\label{sec:intro}

The most fundamental transverse profile for laser beams is the Gaussian profile, where the transverse profile has an amplitude described by the Gaussian function with a characteristic width, and the amplitude profile upon propagation remains a Gaussian that changes in width and acquires spatial phase. In free-space, but also for laser resonators and cavities, the Gaussian beam is therefore a natural lowest-order mode. Many real systems produce beams having a profile very close to a Gaussian such that it is a satisfactory approximation and can be used for modeling propagation and the interaction with matter. However, for high-power lasers, the limited aperture of amplifier crystals and the need for efficiency in pump laser operation and subsequent seed laser amplification, the transverse profile is generally something significantly different from a Gaussian that has a more flat-top character. This results in differences in propagation and focusing, that can have a large effect on certain sensitive applications.

\begin{figure}[htb]
	\centering
	\includegraphics[width=86mm]{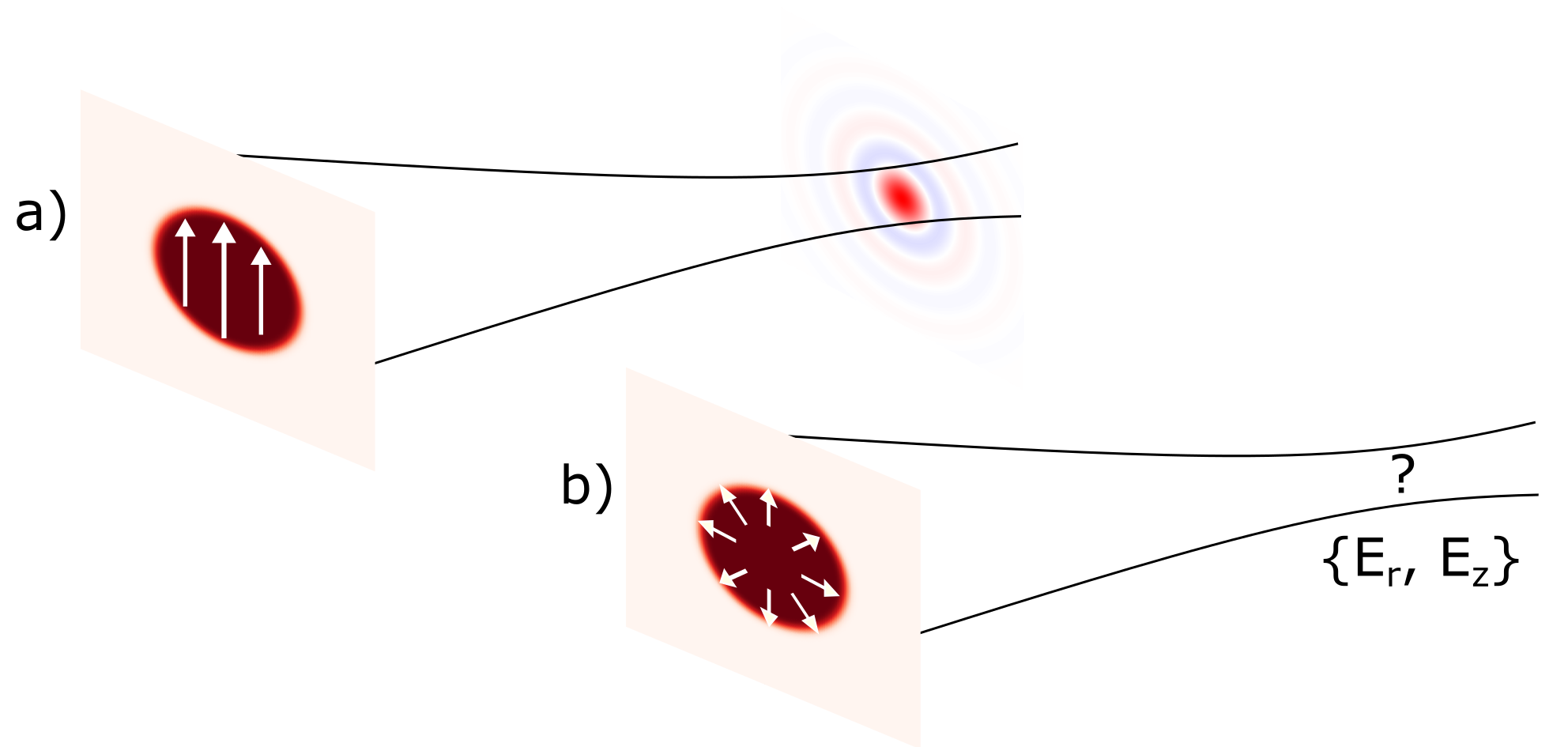}
	\caption{Sketch of the scenario considered. A flat-top linearly-polarized beam (a) creates a more complex radial profile when focused. A flat-top radially-polarized beam (b) should produce both radial and longitudinal fields in the focus, of unknown character. Note that in (b) there will be some region of either linear or undefined polarization near $r=0$, depending on the polarization conversion method used.}
	\label{fig:schematic}
\end{figure}

For scalar laser beams, there are some well-known methods to handle the propagation or focusing of flat-top beams. The flattened Gaussian (FG) beam is a analytical construction meant to represent a linearly-polarized beam approaching a flat-top, Fig.~\ref{fig:schematic}(a), that has analytical solutions to the propagation in free-space~\cite{gori94,bagini96}. The FG is essentially an equivalent to the super-Gaussian~\cite{santarsiero99}, generally used to describe the true near-field profile of most high-energy laser beams, where the super-Gaussian importantly does not have known analytical solutions for its propagation but is more practical to use for the representation of the amplitude at a single plane. Although, as we will see, the FG requires many terms in a sum to describe beams that are very flat-top and have a sharp amplitude cutoff, this description can still be preferable to numerical propagation via a two-dimensional Fourier-transform when demands on the spatial resolution or accuracy of the propagation calculation are high. The analytical forms allow for computational speed-up and can also provide for intuition about general or specific properties of behavior during propagation.

Beyond scalar beams there are many types of vector beams (spatially-varying polarization) that have properties that are fundamentally interesting and useful for some applications. A subset of vector beams are the cylindrical vector beams~\cite{zhanQ09} (CVBs) that have no dependence on the polar angle and only on the radial coordinate, of which the simplest in terms of polarization structure is the radially-polarized (RP) beams. RP beams have been described, generated, and used for various purposes, both in the case of the lowest-order fundamental RP beam~\cite{zhanQ04,kozawa05,kozawa10-2,kozawa18} and higher-order RP beams~\cite{tovar98-2,kozawa06,kozawa07,kozawa10-1,kozawa11,kozawa12,jolly22} (of higher radial order). One of the most well-known advantages or RP beams is that at the focus, due to the local development of a longitudinal field, the width of the intensity profile is below the common diffraction limit~\cite{quabis00,dorn03,kozawa21}. However, the propagation characteristics of RP beams with an arbitrary amplitude profile is not so often studied, mostly because the basics of vector diffraction theory make the problem more complicated.

In this work we consider the case of RP beams that have a flat-top intensity profile before being focused. A schematic for this can be seen in Fig.~\ref{fig:schematic}(b). We are motivated to study this in the context of high-power lasers that inherently have a flat-top profile being converted into RP and then being focused. In this scenario, immediately after the polarization conversion, the beam would truly be a flat-top RP beam, where only at $r=0$ would there be a small area where the polarization would be unknown or would stay linear, related to the limitations of the polarization conversion method in use. Therefore, immediately after the polarization conversion, there would be no intensity zero and true polarization singularity as is commonly associated with RP beams. However, of course, the intensity near $r=0$ would either quickly diffract away on propagation or, if the beam is immediately focused, would contribute a negligible amount to the field at the focus due to it's high associated spatial frequency. No decomposition into analytical RP beams will be able to model a non-zero intensity at the origin, but due to the previously stated diffraction effects in the real-world scenario, we accept that limitation. What we will present in this work is an attempt to model as much as possible the flat-top character of such a beam with RP, which means especially the intensity content close to $r=0$ and the sharp intensity transition at the edges of the beam profile.

In the case of some strong-field laser-matter interactions, modeling the focusing of such a high-power flat-top RP beam with some equivalent low-order RP beam will not be sufficient to accurately predict the dynamics. Conversely, experimentally converting the flat-top to a profile more easy to model when having RP will result in significant losses or additional complexity, potentially spoiling the desired strong-field interaction. An additional application of RP beams with an ultrashort pulse duration is electron acceleration taking advantage of the strong longitudinal field, especially when very tightly focused~\cite{varin05,salamin07,fortin10,wong10,payeur12,carbajo16,wong17-3,powell23}. Most of the modeling of this process has relied on the fundamental RP beam (whether paraxial, with non-paraxial corrections, or as part of exact non-paraxial solutions), which would not include the effects of the initially flat-top profile of the highest power laser beams currently in operation. Therefore an analytical description is desirable. This work has implications as well for the focusing of a flat-top RP beam of any power, since the equations will be based purely on linear optical propagation.

We will first attempt to model a flat-top RP beam with a sum of fundamental beams, and then we will achieve a model analogous to the FG that is a sum of RP beams of many orders. This will be essentially a radially-polarized flattened-Gaussian (RPFG). We will analyze the found solutions for the RPFG, compare the results to integral solutions that we solve numerically, and finally discuss the potential for non-paraxial solutions for the same scenario.

\section{Using a sum of two fundamental beams with different sizes}
\label{sec:sum}

The fundamental radially-polarized Gaussian beam is

\begin{align}
	\begin{split}
		E_r=&E_0 Q^2\rho e^{-Q\rho^2}e^{i\eta} \label{eq:RPLB_Er}
	\end{split}\\
	\begin{split}
		E_z=&-E_0 i\epsilon Q^{2}\left[1-\rho^2{Q}\right]e^{-Q\rho^2}e^{i\eta}, \label{eq:RPLB_Ez}
	\end{split}
\end{align}

\noindent where the normalized radius and axial coordinate are $\rho=r/w_0$ and $\zeta=z/z_R$, $w_0$ and $z_R=kw_0^2/2$ characterize the beam waist and the associated Rayleigh range, $\eta=\omega{t}-kz$ and $k=\omega/c$ for $c$ the speed of light, and $Q=i/(i+\zeta)$. We choose $z=0$ to correspond to the focal area since this is where we are interested in knowing the electric field profile, such that the near-field (before the focusing element) is a some $z\gg z_R$.

We can use intuition from scalar beams, for example the cosh-Gaussian beam, that a superposition of two fundamental RP beams with different beam waists may provide a more flat-top profile while perfectly conserving the radial polarization. We can construct it simply at the focus by the following

\begin{align}
	\begin{split}
		E_r=&\frac{E_0}{\sqrt{1+R^2}} \left[Q_1^2\rho_1 e^{-Q_1\rho_1^2}+RQ_2^2\rho_2 e^{-Q_2\rho_2^2}\right]e^{i\eta} \label{eq:RPLB_Er_sum}
	\end{split}\\
	\begin{split}
		E_z=&\frac{-iE_0}{\sqrt{1+R^2}}\Big\{\epsilon_1 Q_1^{2}\left[1-\rho_1^2{Q_1}\right]e^{-Q_1\rho_1^2} \\
		&+R\epsilon_2 Q_2^{2}\left[1-\rho_2^2{Q_2}\right]e^{-Q_2\rho_2^2}\Big\}e^{i\eta}. \label{eq:RPLB_Ez_sum}
	\end{split}
\end{align}

\noindent In this superposition there are now two beam sizes $w_0^{(1)}$ and $w_0^{(2)}$ (and accordingly $z_R^{(1)}$ and $z_R^{(2)}$) and $Q_i=i/(i+z/z_R^{(i)})$. After propagation away from the focus this forms a beam with a quasi-flat-top, where the $R$ parameter was necessary to decrease the energy of the larger beam at focus such that out of focus it has a peak amplitude similar to the fundamental beam with $w_0^{(1)}$.

\begin{figure}[htb]
	\centering
	\includegraphics[width=86mm]{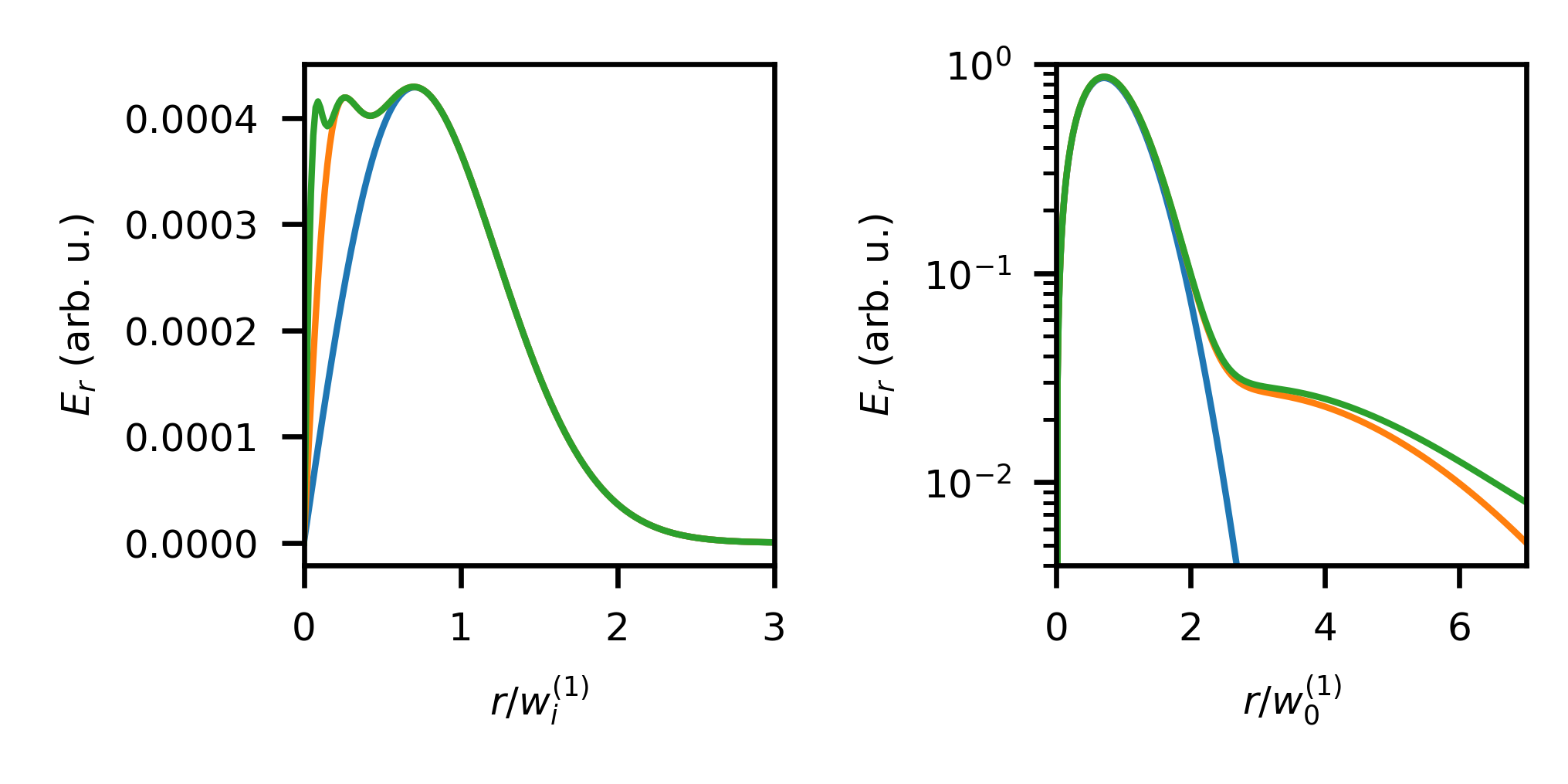}
	\caption{Near-field (left) and far-field (right) of $E_r$ with a single beam (blue), a second beam $R=0.03125$ and $w_0^{(2)}=4w_0^{(1)}$ (orange), and even a third beam with $R=3.472e-3$ and $w_0^{(2)}=12w_0^{(1)}$ (green).}
	\label{fig:sum}
\end{figure}

This strategy can make something that looks generally more flat-top, seen in Fig.~\ref{fig:sum}. And indeed, the feature near $r=0$ in the near-field is relatively sharp with just two beams ($w_0^{(2)}=4w_0^{(1)}$) and could become even sharper with the addition of more beams in the sum with larger (smaller) sizes in the far-field (near-field). This is shown with a third beam ($w_0^{(3)}=12w_0^{(1)}$) also in Fig.~\ref{fig:sum}. However, the features at the outer edge of the beam are still not sharp at all, since every beam in the sum is a fundamental beam and has smooth features away from $r=0$.

The far-field of the sum, seen in the right of Fig.~\ref{fig:sum}, has very little differences from a fundamental beam with $w_0^{(1)}$. We need a log-scale to see deviations from the fundamental RPLB in the far-field. We note that since the transverse field is almost indistinguishable near $r=0$ in the far-field, that the longitudinal field is also barely modified for the three case shown in Fig.~\ref{fig:sum} and is therefore not shown. This is because the second beam in the sum focuses to a much larger size and has a much smaller contribution ($R=0.03125$), and the third beam even more so. Put another way, to have significant differences in the far-field there should be sharp features away from $r=0$ in the near-field, that do not simply diffract away at the focal plane. To have such sharp features in the near-field, higher-order RP beams are necessary. This is what we consider in the following section.

\section{Using an analogue to a flattened Gaussian beam}

In contrast to the previous section where a more flat-top-like beam was created with a sum of lowest-order RPLB beams having different widths, we could attempt to create a flat-top beam using a sum of many orders of RPLB beams with the same width. This is the strategy used for the flattened-Gaussian (FG) beam~\cite{gori94,bagini96}, and is also related to more recent work studying Laguerre-Gaussian (LG) beams as orthogonal bases~\cite{vallone17}, which we will first describe for linear polarization (LP) and then derive for radial-polarization (RP).

\subsection{Linearly-polarized flattened Gaussian}
\label{sec:linear_polarization}

\begin{figure}[htb]
	\centering
	\includegraphics[width=86mm]{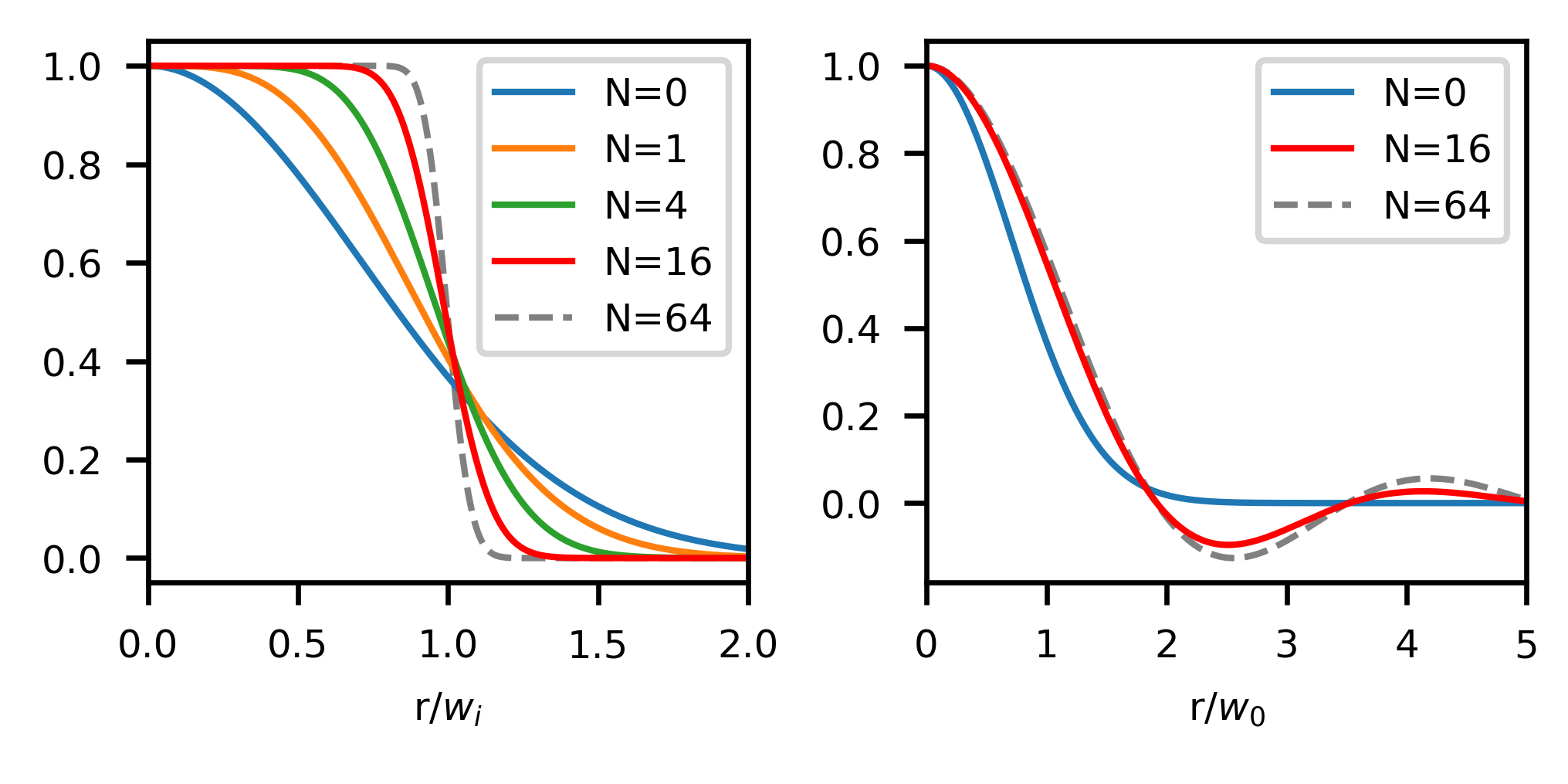}
	\caption{Linearly-polarized flattened Gaussian beam in the nearfield (a) and the farfield (b).}
	\label{fig:FG}
\end{figure}

The FG beam of order $N$ in the nearfield takes the form of

\begin{equation}
	U^{(N)}(r) = \exp{\left(-\frac{(N+1)r^2}{w_i^2}\right)} \sum_{n=0}^{N} \frac{1}{n!} \left(\frac{\sqrt{N+1}r}{w_i}\right)^{2n} \label{eq:FG_NF_sum},
\end{equation}

\noindent where $w_i$ is the characteristic width in the near-field, and the magnitude of the field is normalized to one. With $N=0$ this is a Gaussian beam and as $N$ approaches infinity this approaches $\textrm{circ}(r/w_i)$. This can be equivalently expressed as a sum of weighted LG beams~\cite{gori94}

\begin{align}
	U^{(N)}(r) &= \exp{\left(-\frac{(N+1)r^2}{w_i^2}\right)} \sum_{n=0}^{N} c^{(N)}_n L_n\left(\frac{2(N+1)r^2}{w_i^2}\right) \label{eq:FG_NF_LG} \\
	c^{(N)}_n &= (-1)^n \sum_{m=n}^{N} \frac{1}{2^m}{m \choose n},
\end{align}

\noindent where $L_n$ is the Laguerre polynomial of order $n$. This method creates a more tractable representation of the nearfield of the FG while the difficulty is in calculating the coefficients $c^{(N)}_n$. This representation makes the differences between the FG and the sum from the previous section---the FG in the near-field is a sum of many orders of LG beams that all share the same smaller characteristic width. The smaller characteristic width $\propto w_i/\sqrt{N+1}$ allows the low-order LG beams of the sum to create the flat-top near $r=0$ and the higher-order beams to create the sharp amplitude feature at $r=w_i$. This near-field construction of $U^{(N)}$ is shown in Fig.~\ref{fig:FG}(a).

When considering a near-field FG beam that will be focused for the relevant applications, the expression for the beam profile in the far-field is more important. For an FG of order $N$ this can be shown to be~\cite{bagini96}

\begin{align}
	\begin{split}
		\tilde{U}^{(N)}(\rho_N) &= \frac{1}{N+1}\exp{\left(-\rho_N^2\right)} \sum_{n=0}^{N} L_n\left(\rho_N^2\right)  \\
		&= \frac{1}{N+1}\exp{\left(-\rho_N^2\right)} L^{1}_N\left(\rho_N^2\right),
	\end{split}\label{eq:FG_FF_sum}
\end{align}

\noindent which is the Hankel transform of Eq.~(\ref{eq:FG_NF_sum}) with the proper scaling. In this case $\rho_N=r/w_0\sqrt{N+1}$, $L^{1}_N$ is the generalized Laguerre polynomial with indices $N$ and 1, and $w_0$ is the characteristic width in the far-field, related to $w_i$ by the focal length $f$ and frequency $\omega$ as $w_0=2cf/w_i\omega$. Note that now in the far-field the characteristic width of the sum is $\propto w_0\sqrt{N+1}$, i.e. larger than $w_0$, the converse to the near-field case. The magnitude of the field amplitude is once again normalized to one for simplicity. As with the near-field distribution, with $N=0$ this is a Gaussian beam, but as $N$ approaches infinity this approaches $J_1(2r/w_0)/(r/w_0)$. Eq.~(\ref{eq:FG_FF_sum}) is the solution in the farfield with linear polarization including only paraxial terms. The far-field solution $\tilde{U}^{(N)}$ for various orders $N$ is shown in Fig.~\ref{fig:FG}(b).

The importance of this analysis, specifically Eq.~\ref{eq:FG_NF_LG} and Eq.~\ref{eq:FG_FF_sum}, is that these representations as sums of Laguerre polynomials allow for the propagation of the NF and FF of the FG to be analytically calculated using the known propagation characteristics of Laguerre Gaussian beams. However, crucially, the sum for the near-field in Eq.~(\ref{eq:FG_NF_LG}) involves a factor of 2 in the Laguerre polynomial, where the sum for the far-field in Eq.~(\ref{eq:FG_FF_sum}) does not have a factor of 2. This implies the standard Laguerre Gaussian beam propagation is relevant for the near-field, where the elegant-Laguerre Gaussian propagation is relevant in the far-field. It is the farfield of the FG beam having radial polarization that we calculate in the next section, which involves the radially-polarized elegant-Laguerre Gaussian beam.

\subsection{Radially-polarized analogue to a flattened-Gaussian}
\label{sec:RPFG}

As we saw in the previous section, a linearly-polarized FG beam can be constructed using various sums. For us, the most interesting is the sum in Eq.~(\ref{eq:FG_FF_sum}) that does not involve complicated weighting coefficients, and involved a sum of Laguerre polynomials. To produce the radially-polarized flattened-Gaussian (RPFG) we use the RP analogue to the eLG beam (RPeLG) and use a similar sum to produce the far-field of a proposed RPFG. Then we propagate that back to the near-field (i.e. before focusing) to confirm that it is indeed an RPFG. In fact, we attempted to produce a flat-top-like intensity profile directly in the near-field using a sum of higher-order RPeLG beams using the procedure detailed briefly in Ref.~\cite{jolly22}, but were not successful in producing analytical descriptions for the coefficients. Therefore we present rather the ansatz created first in the far-field (focus) and the confirmation of it's success upon propagation to the near-field.

The paraxial fields of the RPeLG are

\begin{align}
	\begin{split}
		E_{r}^{(n)}=&E_0 e^{i\eta}e^{-Q\rho^2}Q^{n+2}\rho L_{n}^{(1)}(Q\rho^2) \label{eq:RPeLG_Er}
	\end{split}\\
	\begin{split}
		E_{z}^{(n)}=&iE_0 e^{i\eta}e^{-Q\rho^2}Q^{n+2}\epsilon(n+1)L_{n+1}(Q\rho^2), \label{eq:RPeLG_Ez}
	\end{split}
\end{align}

\noindent which at $z=0$ are simply

\begin{align}
	\begin{split}
		E_{r}^{(n)}(z=0)=&E_0 e^{i\omega{t}}e^{-\rho^2}\rho L_{n}^{(1)}(\rho^2) \label{eq:RPeLG_Er_zeta0}
	\end{split}\\
	\begin{split}
		E_{z}^{(n)}(z=0)=&iE_0 e^{i\omega{t}}e^{-\rho^2}\epsilon(n+1)L_{n+1}(\rho^2). \label{eq:RPeLG_Ez_zeta0}
	\end{split}
\end{align}

\noindent One can note the similarity of the radial component of the RPeLG in Eq.~\ref{eq:RPeLG_Er_zeta0} to the terms in the sum form of the farfield of the linearly-polarized FG in Eq.~\ref{eq:FG_FF_sum}. Therefore, the RPeLG could be used as a basis to form a sum that creates a radially-polarized flattened-Gaussian (RPFG) analogue.

We use the RPeLG fields to construct a radially-polarized version of the farfield of the standard FG as in Eq.~\ref{eq:FG_FF_sum} using scaled versions of the RPeLG equations using $\rho_N=r/(w_0\sqrt{N+1})$ as in the linearly-polarized case as an ansatz, and eventually modifying the sum slightly. The scaled versions of the fields are

\begin{align}
	\begin{split}
		E_{r}^{(n,N)}=&E_0 e^{i\eta}e^{-Q_N\rho_N^2}Q_N^{n+2}\rho_N L_{n}^{(1)}(Q_N\rho_N^2) \label{eq:RPeLG_Er_N}
	\end{split}\\
	\begin{split}
		E_{z}^{(n,N)}=&iE_0 e^{i\eta}e^{-Q_N\rho_N^2}Q_N^{n+2}\epsilon_N(n+1)L_{n+1}(Q_N\rho_N^2). \label{eq:RPeLG_Ez_N}
	\end{split}
\end{align}

\noindent where $Q_N=i/(i+z/[z_R(N+1)])$ and $\epsilon_N=w_0/(z_R\sqrt{N+1})$. Combining the model of the FG in focus and the paraxial fields of RPeLG, we can construct the fields of an RPFG of order $N$ as follows

\begin{align}
	E_{r}^{(N)}&=\frac{1}{N+1}\sum_{n=0}^{N} \frac{E_{r}^{(n,N)}}{\sqrt{n+1}}  \\
	E_{z}^{(N)}&=\frac{1}{N+1}\sum_{n=0}^{N} \frac{E_{z}^{(n,N)}}{\sqrt{n+1}}  ,
\end{align}

\noindent where the sum is only changed by the factor of $1/\sqrt{n+1}$ from that for the linearly-polarized flattened-Gaussian beam, Eq.~\ref{eq:FG_FF_sum}, and there is no need to calculate any coefficients as for Eq.~\ref{eq:FG_NF_LG}. These sums do not admit simple solutions, but at the very least do not require numerical solutions like integrals do. Even the most simple sum, $E_{z}^{(N)}(r=0)\propto Q_N^2\sum_{n=0}^{N} Q_N^n\sqrt{n+1}/(N+1)$ does not have a simpler solution except for exotic functions that are themselves sums (polylogarithmic functions and the Lerch transcendent).

\begin{figure}[htb]
	\centering
	\includegraphics[width=86mm]{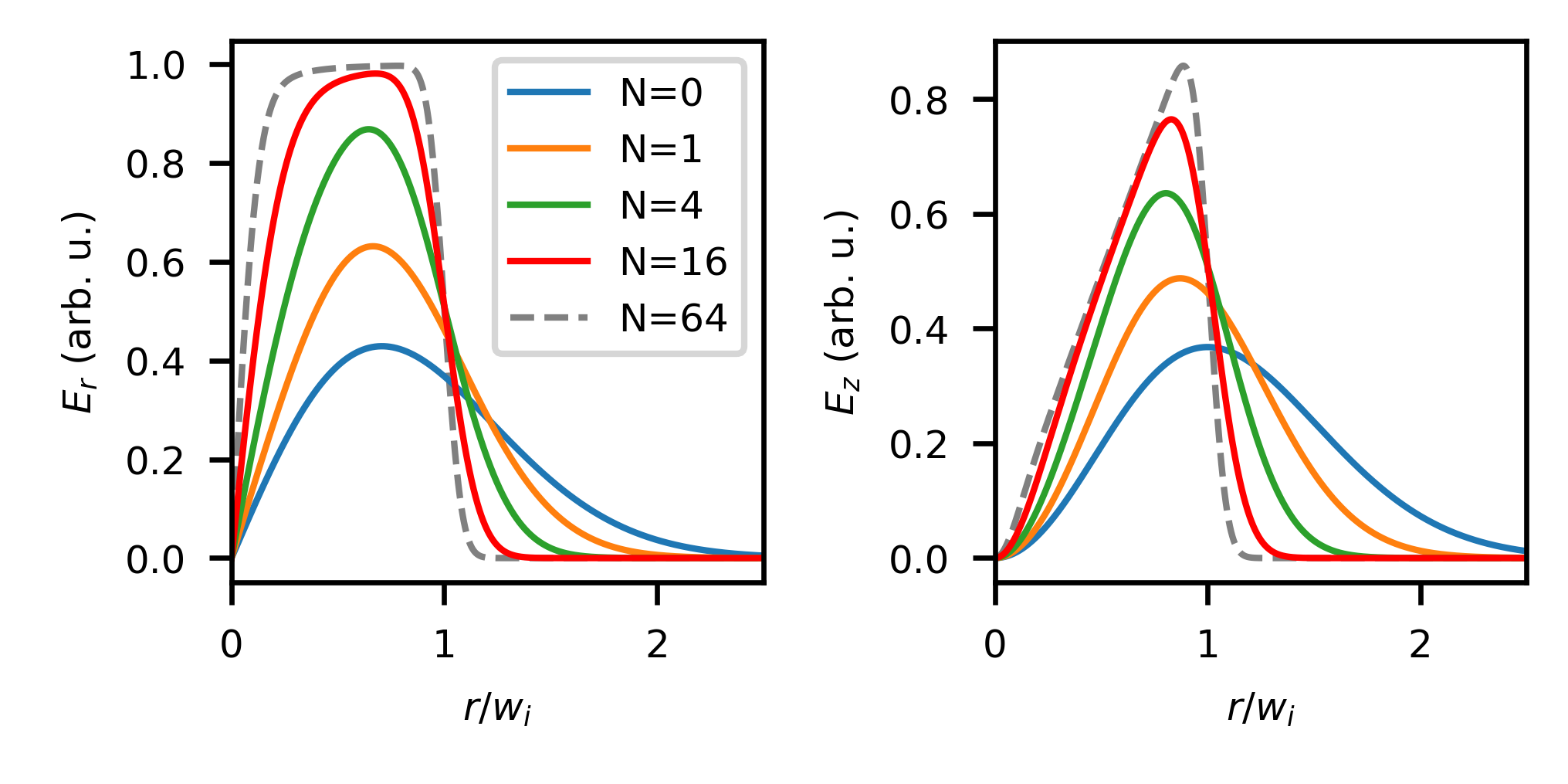}
	\caption{Radially-polarized flattened Gaussian (RPFG) beam in the near-field ($z\gg z_R$) for various values of $N$. The transverse field $E_r$ (a) becomes more flat-top, while the longitudinal field (b) is linear in $r$.}
	\label{fig:RPFG_NF}
\end{figure}

Somewhat counter-intuitively we have so far constructed the RPFG at its focus around $z=0$ using an ansatz, i.e. at the far-field. Therefore it must be checked that indeed at large $z\gg z_R$ the amplitude profile approaches a flat-top. This can be seen in Fig.~\ref{fig:RPFG_NF} for both $E_r$ and $E_z$ for a number of $N$ values, confirming the flat-top character. Note that with increasing $N$ the intensity transition at $w_i$ becomes steeper and the energy content pushes closer to $r=0$, but all cases still are forced to have zero intensity at $r=0$. Due to the nature of diffraction the longitudinal field becomes small compared to the transverse field upon propagation away from the focus such that $E_r$ is the main contribution to the overall amplitude profile, which we can see is clearly more flat-top as $N$ increases. Due to symmetry considerations as $N$ increases $E_z\propto r$ when $r<w_i$.

\begin{figure}[htb]
	\centering
	\includegraphics[width=86mm]{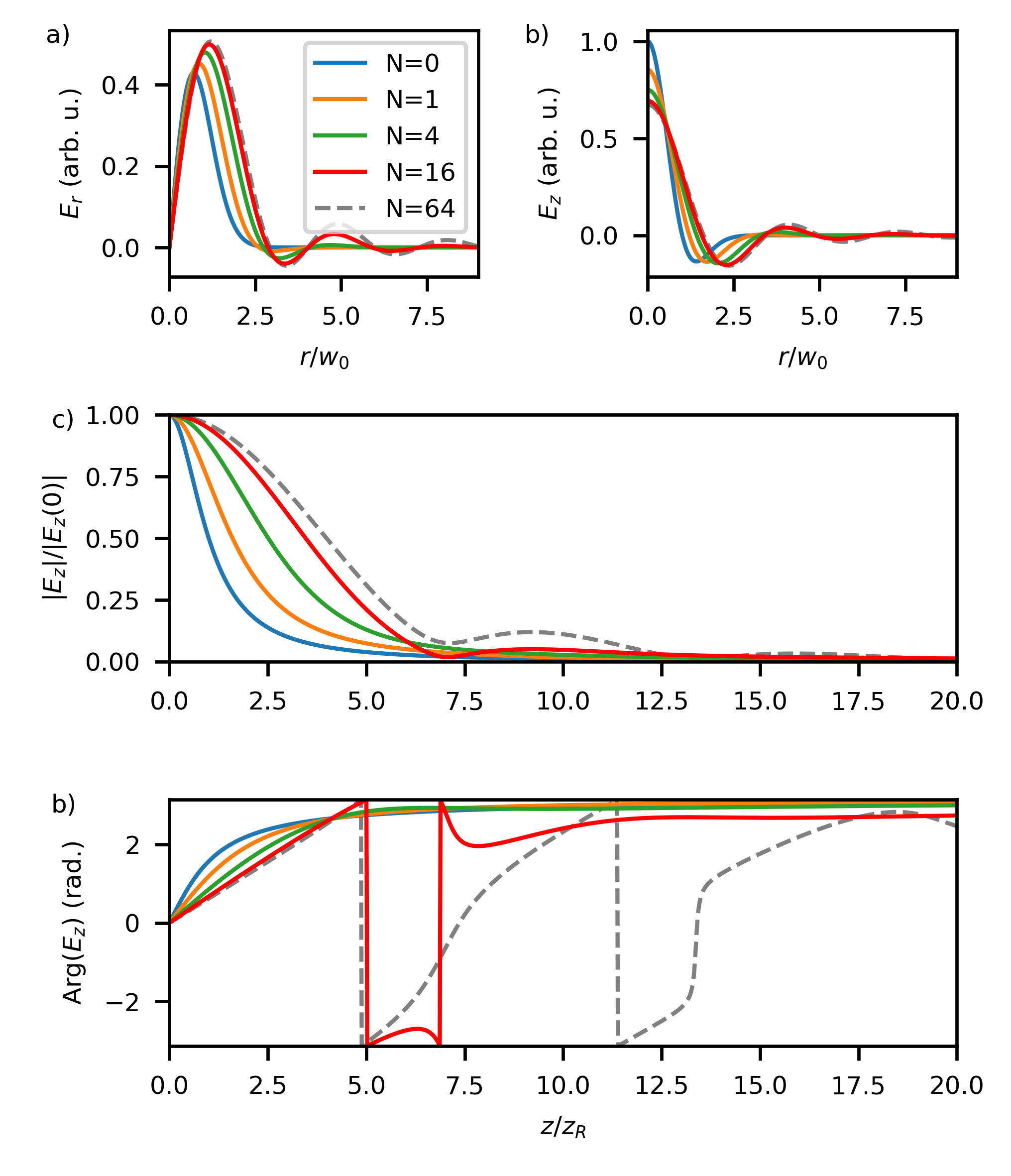}
	\caption{Radially-polarized flattened Gaussian (RPFG) beam in the far-field. The transverse field $E_r$ (a) and longitudinal field $E_z$ (b) are shown for various values of $N$ at $z=0$, along with the amplitude (c) and phase (d) evolution of $E_z$ as $z$ increases (i.e. upon propagation).}
	\label{fig:RPFG_FF}
\end{figure}

The fields at the focus and the behavior of $E_z$ at $r=0$ under propagation can be seen in Fig.~\ref{fig:RPFG_FF}. As is expected from Fourier optics intuition, as $N$ increases, both $E_r$ and $E_z$ become slowly more extended transversely (due to more content near $r=0$ in the near-field) and have more amplitude oscillations outside of the main ring or inner lobe, respectively (due to the sharper boundary at $r=w_i$ in the near-field). In the focus at $r=0$ the amplitude of $E_z$ decreases more slowly as $N$ increases, and at high enough $N$ begins to show non-monotonic behavior.

It is important to note the phase at $r=0$ for the longitudinal field, Fig.~\ref{fig:RPFG_FF}(d), agrees with past conclusions~\cite{pelchat-voyer20} where, once there is a sharp enough boundary in the illumination, the phase will deviate from the standard total Guoy phase difference from $z=-\infty\rightarrow+\infty$ for an RPLB of $2\pi$. This is seen in the $N=64$ case in Fig.~\ref{fig:RPFG_FF} where from $z=0$ to large $z$ the phase undergoes almost $5\pi$, 5 times the expected amount. This does not present in the $N=16$ case, where the phase briefly exceeds $\pi$ but settles at larger $z$.

Another interesting observation in the phase is that as $N$ becomes larger, the phase (along with the amplitude) changes more slowly with $z$. This was shown to be the case for RPLBs where the derivative of the illumination amplitude at $r=0$ was modified~\cite{pelchat-voyer21}, which turned out to be relevant to direct electron acceleration efficiency with ultrashort RP pulses~\cite{pelchat-voyer22}.

\begin{figure}[htb]
	\centering
	\includegraphics[width=86mm]{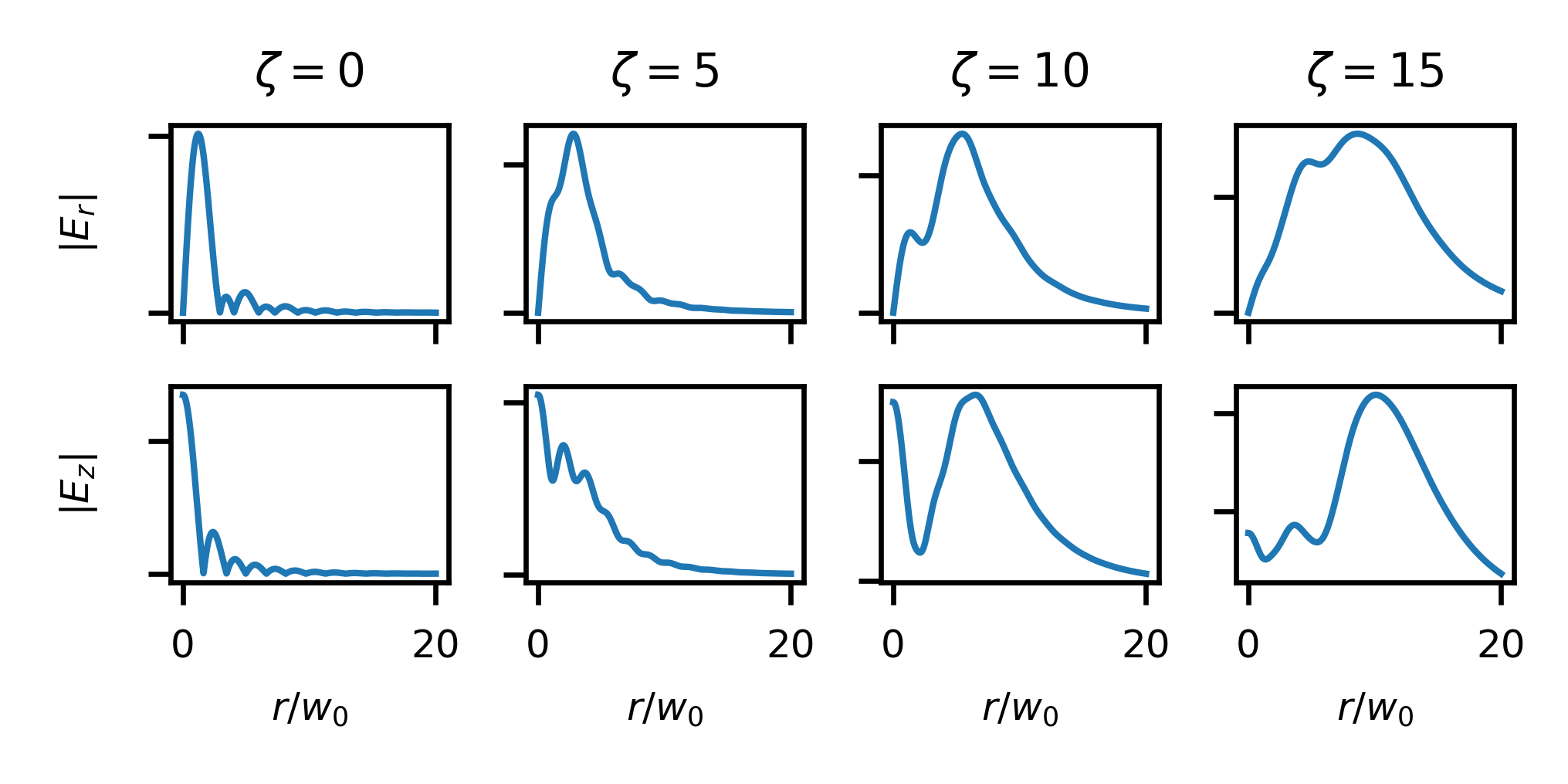}
	\caption{Radially-polarized flattened Gaussian (RPFG) beam propagating out of the far-field for $N=64$, with $|E_r|$ (top) and $|E_z|$ (bottom) for 4 different values of $\zeta=z/z_R$ (left to right).}
	\label{fig:RPFG_FF_prop}
\end{figure}

As a last detail we show in Fig.~\ref{fig:RPFG_FF_prop} the amplitude profile of both components of the field for different positions around the far-field. This shows the evolution from a localized beam to a beam beginning to have flat-top features and a sharp amplitude drop-off, and importantly the complex evolution between those two extremes. The intensity distribution near the focus depends on the strength of focusing, i.e. $\epsilon$ determines the relative strength of $E_z$ to $E_r$, but far away from the focus $E_r$ dominates regardless of the focusing.

For applications it is interesting to know the physical beam power in a certain beam, or to be able to properly describe a beam given a known beam power, where so far we have just modeled the fields with a placeholder constant $E_0$. The power can be calculated via the Poynting vector flux such that the power for a given order $n$ of RPeLG is

\begin{align}
	P^{(n)} = \varepsilon_0 c \pi E_0^2 w_0^2 \int_{0}^{\infty}\rho^3e^{-2\rho^2}[L_n^1(\rho^2)]^2\,d\rho. \label{eq:RPeLG_power}
\end{align}

\noindent Then the power for an RPFG of order $N$ is

\begin{align}
	P^{(N)} = \frac{\varepsilon_0 c \pi E_0^2 w_0^2}{N+1} \sum_{n=0}^{N}\frac{\int_{0}^{\infty}\rho^3e^{-2\rho^2}[L_n^1(\rho^2)]^2\,d\rho}{\sqrt{n+1}}. \label{eq:RPFG_power}
\end{align}

\noindent If one fixes the power of an RPFG, then this equation must be inverted to solve for $E_0$.

\section{Comparison to integral solutions}
\label{sec:integral}

There are various integral solutions that could be used to approximate the focusing of a flat-top RPLB or an RPFG. One example is using a vectorial diffraction integral in the Stratton-Chu~\cite{stratton39} or Richards-Wolf~\cite{richards59} formalism. A radially-polarized beam can also be constructed from a sum of circularly polarized vortex beams~\cite{porras21} therefore requiring only a scalar diffraction integral for the radial field and a calculation of the longitudinal field. In this section we compare a simplified integral solution to the solutions based on a sum of Laguerre beams from the previous section.

Assuming cylindrical symmetry and an illumination of $\textrm{circ}(r/w_i)$, i.e. a perfect flat-top illumination, the paraxial approximation allows for simplifying the opening half-angle $\alpha$ such that $\cos(\alpha)\sim1-\alpha^2/2$ and $\sin(\alpha)\sim\alpha$ reducing the integral solution to~\cite{pelchat-voyer21}

\begin{align}
	E_r &= e^{-ikz}\int_{0}^{w_i/f}i\alpha(1-\frac{\alpha^2}{2})J_1(kr\alpha)e^{ikz\alpha^2/2}\,d\alpha\label{eq:E_r_integral}\\
	E_z &= e^{-ikz}\int_{0}^{w_i/f}\alpha^2 J_0(kr\alpha)e^{ikz\alpha^2/2}\,d\alpha\label{eq:E_z_integral}.
\end{align}

\noindent Since this is a finite integral, there is not an analytical solution as far as we know, but it can be integrated numerically.

\begin{figure}[htb]
	\centering
	\includegraphics[width=86mm]{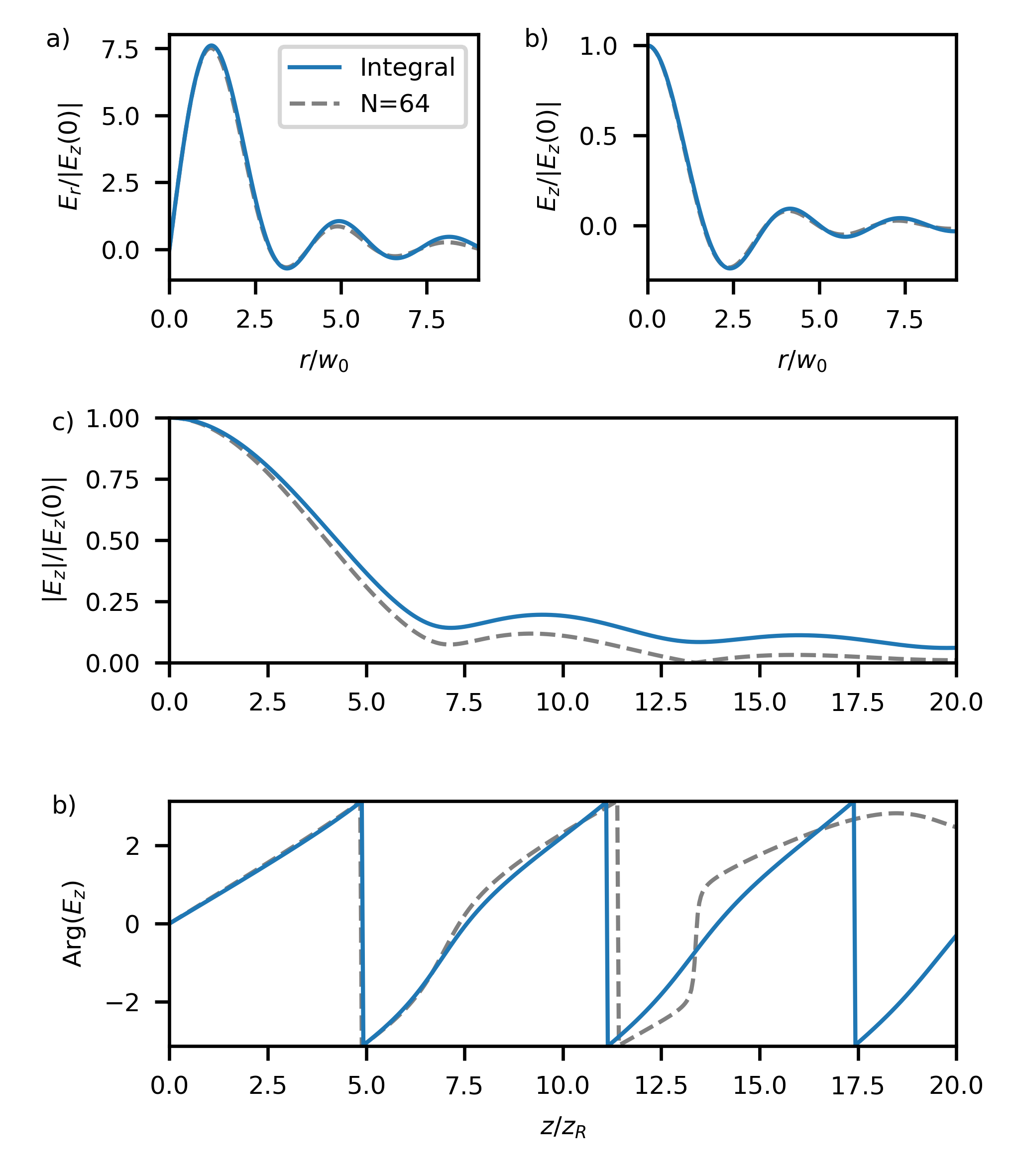}
	\caption{Farfield of the integral solution with a flat-top illumination in the paraxial regime (blue solid line) compared to the RPFG solution for $N=64$ (gray dashed line). To have a quantitative comparison, the parameters were chosen to be: $\lambda_0=800$\,nm, $w_i=5$\,cm, $f=50$\,cm, such that $w_0/z_R=0.1$.}
	\label{fig:RPFG_FF_integral}
\end{figure}

The numerical integration for the paraxial and perfectly flat-top RP beam is shown in Fig.~\ref{fig:RPFG_FF_integral}. When comparing to the $N=64$ case shown as a dashed line, it is clear that the perfectly flat-top case is in agreement with the characteristic features taken to the extreme. The off-axis modulations in $E_r$ and $E_z$ at $z=0$ are qualitatively similar to the $N=64$ RPFG, but slightly larger in relative magnitude. The longitudinal field $E_z$ at $r=0$ experiences the same slower decrease as $z$ increases, also with modulations, again similar but larger in magnitude than for the $N=64$ RPFG. Finally, the on-axis phase of $E_z$ in Fig.~\ref{fig:RPFG_FF_integral} shows similar behavior as in Fig.~\ref{fig:FG}, except that it continues to increase for a greater range of $z$ such that it goes through $\sim6\pi$ from $z/z_R=0$ to 20.

The good agreement between the $N=64$ RPFG (which is analytical albeit requiring many terms in a sum) and the numerical solution to the vector diffraction integral, is also a sign that the energy content near $r=0$ is not a dominating factor. Even the high-order RPFG has zero field at $r=0$ in the near-field (see Fig.~\ref{fig:RPFG_NF}), but the illumination of the vector diffraction integral Eqs.~\ref{eq:E_r_integral}--\ref{eq:E_z_integral} is an exact flat-top (rigorously with a punctual and therefore arbitrarily small hole at $r=0$). Still, they agree very well in-focus in Fig.~\ref{fig:RPFG_FF_integral}. The disagreement at moderate $z/z_R$ values, i.e. along propagation, is more likely due to the integral solution being for a perfectly sharp flat-top illumination, where even the $N=64$ RPFG has smooth features (see again Fig.~\ref{fig:RPFG_NF}).

\section{Non-paraxial corrections}

Non-paraxial corrections are interesting, because the longitudinal field is the strongest with tighter focusing, and therefore the width of the beam intensity reaches a much smaller level. However, of course, non-paraxial models generally require increased complexity.

Intermediate non-paraxial corrections can be based on the model in Ref.~\cite{yan07} which is conveniently based on the RPeLG in the non-paraxial regime. This would involve taking the non-paraxial corrections for $E_r$ and $E_z$ in Ref.~\cite{yan07} and apply them to the the RPeLG fields at each value of $n$, and then do the same sum to calculate the RPFG for a given $N$. However, the non-paraxial corrections at each $n$ cause a spurious increase in the total energy that depends on $n$ such that higher-order parts of the sum would be over-represented. In the sum over $n$ we would need to compensate for this, but since the over-contribution is not known analytically (i.e. not derived in Ref.~\cite{yan07}), this is not currently tractable.

Arbitrarily non-paraxial field equations could be based on the model of April~\cite{april08-1,april08-2}. Generally, the arbitrarily non-paraxial model for the RPeLG at a given $n$ requires itself a sum over a higher-order vector-potential solution to the Helmholtz equation involving the spherical Bessel function and the Legendre function. Then, to construct the arbitrarily non-paraxial RPFG a sum of these non-paraxial RPeLG would be performed over $n$ to $N$ and derivatives taken to calculate the fields from the vector potential. Evidently this is very complicated and we consider it outside the scope of this work.

Finally, integral models used in Section~\ref{sec:integral} have non-paraxial forms. These involve un-simplified $\sin(\alpha)$ and $\cos(\alpha)$ terms, and require a so-called apodization function that depends on the focusing element. The non-paraxial model is useful when $\alpha_\textrm{max}$ (which was implicitly approximated to be $w_i/f$ in the paraxial case) becomes large. The non-paraxial version of Eqs.\ref{eq:E_r_integral}--\ref{eq:E_z_integral} could be reproduced here, but they are outside of the interest of this work since we aim to model the focusing of a flat-top RP beam analytically and we don't yet have an non-paraxial analytical model to compare to the non-paraxial integral model.

\section{Conclusion}
\label{sec:conclusion}

In this work we have shown how the focusing of a radially-polarized (RP) light beam can be modeled using a sum of transversely scaled RP elegant Laguerre-Gaussian (RPeLG) beams, which therefore allows for the propagation of the complex field to be described fully analytically. This required a sum of higher-radial-order RPeLG beams, where the higher the maximum mode $N$ allows for more closely reproducing the flat-top intensity profile. In parallel to this analytical description, we could describe the evolution of the on-axis longitudinal electric field around the focus, since this field is often interesting for applications. The closer to a flat-top before focusing, the more gradually the intensity and phase evolve through the focus, although this creates more complexities away from the optical axis.

The descriptions we have found in this work and the technique we have employed to model an RP flattened-Gaussian analogue may be useful for describing high-intensity laser-matter interactions where a high-power laser with a flat-top intensity is converted to RP. Additionally, it may be possible to use the same technique, a sum or transversely-scaled higher-order RP beams, to describe analytically RP beams with different or more arbitrary transverse intensity profiles. 
	
\section*{Funding}
Fonds De La Recherche Scientifique - FNRS; Horizon 2020 Framework Programme (801505)
	
\section*{Acknowledgments}
S.W.J. has received funding from the European Union’s Horizon 2020 research and innovation program under the Marie Skłodowska-Curie grant agreement No 801505.
	
\section*{Disclosures}
The author declares no conflicts of interest.
	
\section*{Data availability}
Data underlying the results presented in this paper are not publicly available at this time but may be obtained from the authors upon reasonable request.

\end{document}